\documentstyle[lprocl,epsfig,11pt]{article}

\def\Journal#1#2#3#4{{#1} {\bf #2}, #3 (#4)}

\def\NPB{{\em Nucl. Phys.} B}
\def\PLB{{\em Phys. Lett.}  B}
\def\PRL{\em Phys. Rev. Lett.}
\def\PRD{{\em Phys. Rev.} D}
\def\ZPC{{\em Z. Phys.} C}

\def\be{\begin{equation}}
\def\ee{\end{equation}}
\def\bea{\begin{eqnarray}}
\def\eea{\end{eqnarray}}

\begin{document}
\hfill CLNS 97/1521
\title{Heavy Quark Decays\footnote{Invited Talk at the
XVIII International Symposium on Lepton-Photon Interactions}}

\author{ PERSIS S. DRELL }

\address{Cornell University, Newman Laboratory  \\ Ithaca,
NY 14853-5001}


\maketitle\abstracts{
 }

\section{Introduction}

The field of heavy quark decays was born in 1976 with the discovery of the 
$D$ meson at SPEAR \cite{spear}, and has blossomed ever since. By 1981, the 
weak decays of heavy quarks merited their own byline at Lepton--Photon in 
Bonn ~\cite{bonn} and that tradition has continued to the present day.

The new results on heavy quark decays submitted to this year's 
Lepton--Photon Conference naturally fell into two categories around 
which this review is organized.  These are  heavy quark decays at tree level and 
heavy quark decays beyond tree level.

Heavy quark decays at tree level have been studied since 1976, and this is 
indeed a very mature field. These analyses have two goals. The first is the 
study of decay dynamics and the effects of the strong interactions on the 
underlying weak decay. The second goal is to measure the magnitudes of the 
elements of the CKM matrix, $|V_{ub}|$ and $|V_{cb}|$, that represent quark 
flavor mixing ~\cite{ckm}, thereby serving to define the Standard Model.
Rather than  summarizing the status of the tree level decays which has
been done recently in several excellent reviews ~\cite{richman,lkg}, I 
will focus here on several new analyses that are probing the details of 
$b$ heavy quark decays at tree level and testing our understanding
of them. I
will also review charm decays that are aiding us in this effort.  

The second category of heavy quark studies, heavy quark decays beyond tree level,
 represents a new and emerging 
(anything but mature) field. 
Enormous data sets from CLEO and LEP are allowing $b$ quark rare decays to 
be seen and measured for the first time, and there are many new results this 
year. The goals for these studies are either to uncover physics beyond the Standard Model 
or to probe the phases of the elements of 
the CKM matrix.

\section{Heavy Quarks at Tree Level}

\subsection{Semileptonic $B$ to Charm Decays and $|V_{cb}|$}

\begin{figure}
\centerline{\psfig{figure=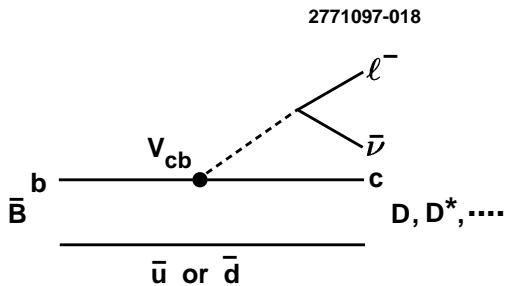,height=1.5in}}
\caption{Diagram illustrating the semileptonic decay of a $\bar B$ meson to a charm
meson.
\label{fig:sldecay}}
\end{figure}

I will start my discussion of tree level heavy quark decays with semileptonic 
decays of the bottom quark as illustrated in
figure ~\ref{fig:sldecay}.
Semileptonic decays are among the most extensively studied of all 
heavy quark decays. This is due to their theoretical simplicity: the matrix 
element can be written as a product of a leptonic current which is exactly 
known, and a hadronic current which can be parameterized in terms of form 
factors. Strong interactions are quite important in these decays, but they are 
sufficiently simple to allow detailed theoretical predictions that can be tested 
experimentally.  These decays are experimentally
accessible with relatively large  branching ratios and clean signatures.

\subsubsection{Exclusive and Inclusive $B$ Semileptonic Branching Ratios}

\begin{figure}
\centerline{\psfig{figure=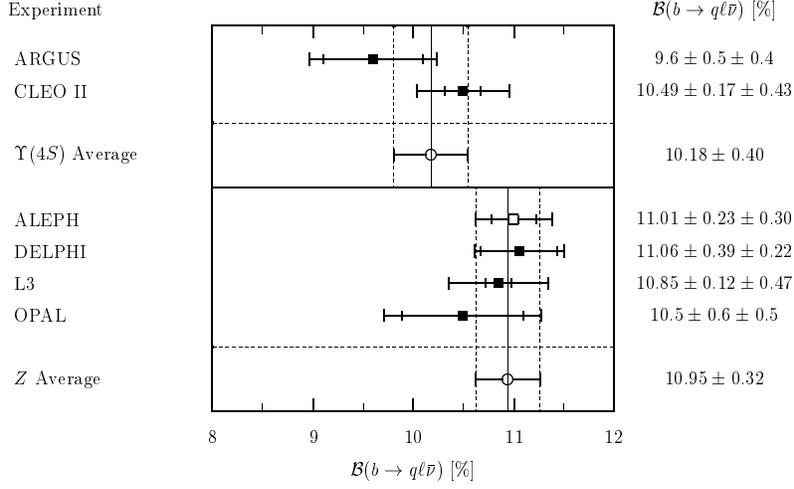,height=2.5in}}
\caption{Summary of the measurements of 
the inclusive semileptonic branching ratio
${\cal B}(b \rightarrow q \ell^-\bar{\nu})$ from 
the $\Upsilon(4S)$ \protect \cite{4s-inc}\protect and
 the $Z$ \protect \cite{z-inc} \protect.
\label{fig:sl1}}
\end{figure}

\begin{figure}
\centerline{\psfig{figure=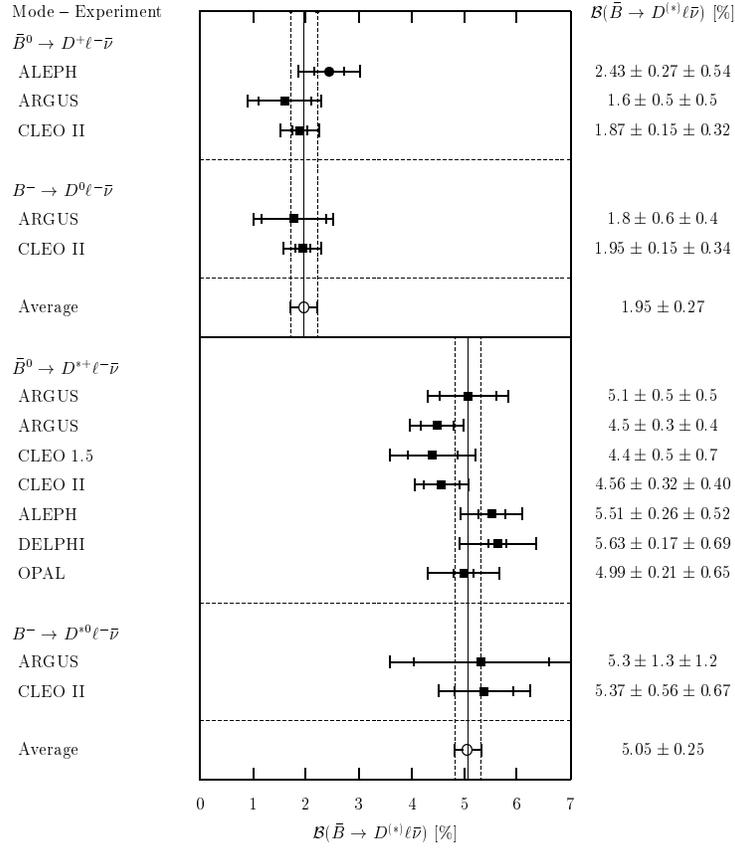,height=4.4in}}
\caption{Summary of measurements of ${\cal B}(\bar B \rightarrow 
D \ell^-\bar{\nu})$ \protect \cite{d-exc} \protect and
${\cal B}(\bar B \rightarrow D^* \ell^-\bar{\nu})$
\protect \cite{ds-exc,ds-exc-vcb} \protect.  In the values listed, the branching
ratios are updated to ${\cal B}(D^0 \rightarrow K^-\pi^+) = 0.0388 \pm 0.0010$
and ${\cal B}(D^+ \rightarrow K^-\pi^+\pi^+) = 0.088 \pm 0.006$
\label{fig:sl2}}
\end{figure}

In table~\ref{tab:slbr}, I give updates on the average inclusive semileptonic
$B$ branching fractions as well as updates on the dominant exclusive semileptonic
branching fractions.  These branching fractions have changed very little in the last
year.  Figures ~\ref{fig:sl1} and ~\ref{fig:sl2} summarize the current experimental
measurements from which the world averages are built.

\begin{table}[t]
\caption{Average inclusive and exclusive semileptonic branching ratios.
\label{tab:slbr}}
\vspace{0.4cm}
\begin{center}
\begin{tabular}{|c|c|}
\hline
 &  \\
 Decay channel & World average branching ratio \\
 &   \\  \hline
 &  \\ 
${\cal B}(\bar B \rightarrow D \ell^-\bar{\nu})$  & $ 0.0195 
\pm 0.0027 $ \\
${\cal B}(\bar B \rightarrow D^* \ell^-\bar{\nu})$  & $ 0.0505 \pm 0.0025 $ \\
${\cal B}(\bar B \rightarrow D^{(*)}\pi X \ell^-\bar{\nu})$  & $ 0.023 \pm 0.0044 $\\
${\cal B}(\bar B \rightarrow X_u \ell^-\bar{\nu})$  & $ 0.0015 \pm 0.001 $\\
 &  \\ 
\hline
 &  \\ 
$\Sigma{\cal B}_i $ &  $0.0945 \pm 0.0058$ \\
${\cal B}(b \rightarrow q \ell^-\bar{\nu})^{\Upsilon(4S)}$  & $ 0.1018 \pm 0.0040 $ \\
${\cal B}(b \rightarrow q \ell^-\bar{\nu})^Z$  & $ 0.1095 \pm 0.0032 $ \\
 &  \\ 
\hline
\end{tabular}
\end{center}
\end{table}

The most basic check of our understanding of semileptonic $B$ decays is to 
see that the sum of the exclusive modes is consistent with the inclusive rate. 
Do we understand all of the pieces of the semileptonic rate? The dominant 
semileptonic modes are reasonably well known; however they only account 
for 70 percent of the total semileptonic rate. The remaining 30 percent of the semileptonic rate 
is semileptonic decay to $P$ wave and higher angular momentum charm states, to
 radially 
excited charm states, or to 
non-resonant $D^{(*)}\pi\ell^-\bar{\nu}$ states.  The 
ALEPH collaboration has used a topological study to estimate the contribution 
to the semileptonic rate from states other than $D\ell^-\bar{\nu}$ and 
$D^*\ell^-\bar{\nu}$ ~\cite{dssaleph}.

\begin{figure}
\centerline{\psfig{figure=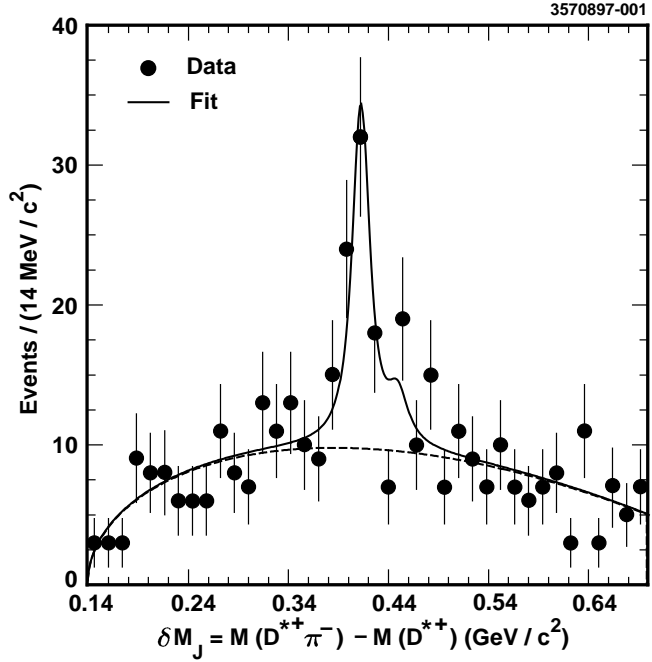,height=3.5in}}
\caption{Observation of semileptonic $B$ decay to a $P$ wave charm meson
by the CLEO detector.  The mass difference  
$M(D^{*+}\pi^-) - M(D^{*+})$
is shown with a clear enhancement at the $D_1$ mass.  
\label{fig:dss}}
\end{figure}

\begin{figure}
\centerline{\psfig{figure=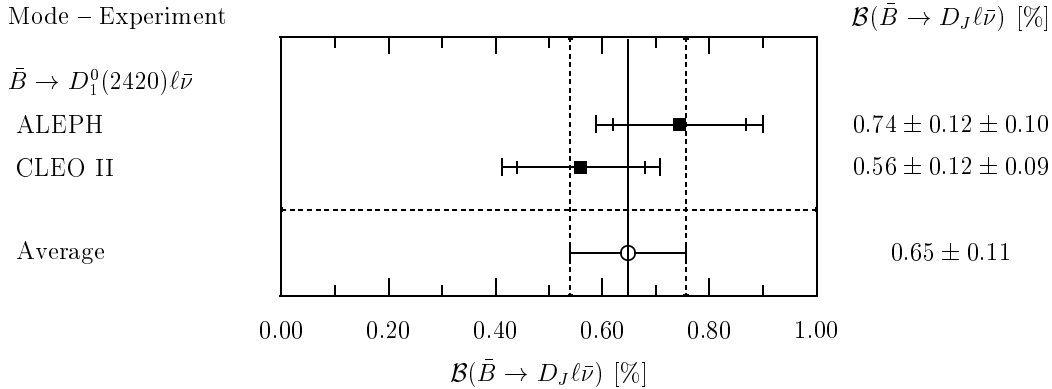,height=2.0in}}
\caption{Summary of the CLEO \protect \cite{cleodss} \protect and
ALEPH \protect \cite{dssaleph} \protect  measurements of ${\cal B}(\bar B \rightarrow D_1 \ell^-\bar{\nu})$. 
\label{fig:dss1}}
\end{figure}

From table~\ref{tab:slbr}, we see that at the $1-2 \sigma$ level, it appears that exclusives saturate the 
inclusives.  However, of the 30\% of the semileptonic 
rate that is not to $D\ell^-\bar{\nu}$ and
$D^*\ell^-\bar{\nu}$,
only one of the decays has been 
exclusively observed. Figure  ~\ref{fig:dss}  shows the new CLEO result for
$\bar B\to D_1\ell^-\bar{\nu}$ ~\cite{cleodss} and figure ~\ref{fig:dss1} 
summarizes the ALEPH and CLEO
measurements
of $\bar B\to D_1\ell^-\bar{\nu}$.  Both experiments also
put upper limits on the amount of $D_2^*$ production in
semileptonic $B$ decays at the 1\% level.
It is puzzling that the $D_1$ and $D_2^*$ can account for so little 
of the total semileptonic
rate, and detailed understanding of the decays other than $D\ell\bar
{\nu}$ 
and $D^*\ell^-\bar{\nu}$ is absent.

\subsubsection{HQET}

Our understanding of how to interpret semileptonic heavy quark decays has 
evolved enormously with the development of heavy quark effective theory 
(HQET)~\cite{hqet}. This allows us to extract CKM matrix
elements from exclusive semileptonic $B$ decays such as 
$\bar B\to D^*\ell^-\bar{\nu}$ with much greater confidence.

The central insight of HQET is to notice that a $B$ meson or a charm meson 
(a light quark bound to a very heavy quark) looks a lot like the hydrogen 
atom  (a light electron bound to a heavy proton). If we
recall that the $e^-$ wave function in hydrogen looks just
 like the $e^-$ wave function in deuterium up to hyperfine
splittings, we expect that 
the light quark part of the B meson wave function should
 look just like the light quark part of the D
meson
wave function, up to hyperfine splittings of order $\Lambda_{QCD}/M_Q$,
where $M_Q$ is the mass of the heavy quark in the meson.

One consequence of this observation is that the
rate for heavy meson to 
heavy meson semileptonic decays
can be factorized into a heavy quark part which is calculable 
and a light quark overlap integral.  This light quark overlap integral 
is not calculable from first 
principles but it is universal. It is the same for all heavy meson to 
heavy  meson decays involving a pseudoscalar or vector meson (up to hyperfine
effects).
All three form factors needed to describe the 
hadronic current in $\bar B\to D^*\ell^-\bar{\nu}$ decay and the one form factor needed
to describe $\bar B\to 
D\ell^-\bar{\nu}$ decay can be written as known quantities times one unknown
function, 
$\xi(v \cdot v')$, where $v$ and $v'$ are the four velocities of
the incoming and outgoing mesons respectively.

\subsubsection{Extraction of $|V_{cb}|$}

A standard method for extracting $|V_{cb}|$ in the past five years has been to 
use exclusive $\bar B\to D^*\ell^-\bar{\nu}$ decays and to take advantage of HQET. 
What makes HQET attractive in analyzing these decays is that at zero recoil, 
when the initial and final state mesons are at rest, the unknown function 
$\xi(v \cdot v')$
 describing the overlap of the initial and final light quark 
wave functions is absolutely normalized. This absolute normalization is a 
result of the fact that in the zero recoil
configuration, the light quark does 
not know or care that a heavy $c$ quark has replaced a heavy $b$ quark
in the decay. 
The overlap is perfect. In the limit of infinitely 
heavy quarks, at this magic kinematic point, $|V_{cb}|$ can be measured 
independent of any unknown form factor.  Effectively one trades statistics
in data to measure $|V_{cb}|$ in a region of phase space where the form factor
is well known.

In practice, the LEP experiments, ARGUS and CLEO have extracted $|V_{cb}|$ from 
$\bar B\to D^*\ell^-\bar{\nu}$ by measuring the differential decay rate as
a function of $w=v \cdot v'= (m^2+M^2-q^2)/(2mM)$, where $M$ and $m$ are
the masses of the initial and final state mesons respectively ~\cite{neubert1}.
In the limit of infinitely heavy quarks, the decay
rate at zero recoil or
 $w = 1$ (corresponding to the
maximum value of $q^2$ for the decay),
where the light quark overlap integral or form factor
is identically one, yields $|V_{cb}|$.

The differential decay rate can be expressed as:
\begin{eqnarray}
{d\Gamma\over dw} & = & {G_F^2\over 48\pi^3} 
\kappa\left(m_B,m_D,w\right) \left|V_{cb}\right|^2{\cal F}^2 
\left(w\right)
\end{eqnarray}
where ${\cal F}(w)=\eta_A \hat{\xi}(w)$, and
$\kappa$ is a known function. In the limit of infinitely heavy quark
masses, $\hat{\xi}(w)$ reduces to  $\xi(w)$.  $\eta_A$ is a correction to the
differential decay rate that can be calculated in perturbative QCD~\cite{czarnecki}.

There are several subtleties which must be kept in mind in this procedure.  The
experiments are now 
sufficiently precise that these corrections are important.

\begin{enumerate}
\item In the real world, $q^2$ or $w$ is not reconstructed 
exactly.  There is experimental 
smearing between the true value and reconstructed value.
An unbinned maximum likelihood fit to the entire $w$
distribution must be done to properly account for the smearing.  
Most experiments now use such a procedure.

\item Also in the real world, the $c$ and $b$
quark masses are not infinitely heavy, and the heavy  quark
symmetry limit is only the first term in an expansion in 
$1/M_{\rm Q}$. The extrapolation to zero recoil gives the form factor
at zero recoil, ${\cal 
F}\left(1\right)$, times $|V_{cb}|$. 
In the limit of infinite quark mass, ${\cal F}\left(1\right)=1$. 
For finite mass quarks the corrections are 
substantial with significant errors ~\cite{ff1calc,neubertrhohat}:
\begin{eqnarray}
{\cal F}\left(1\right)_{D^*\ell\nu}& = & 0.91\pm0.03 \\
{\cal F}\left(1\right)_{D\ell\nu} & =   & 0.98\pm0.07
\end{eqnarray}
\item The shape of the form factor
 distribution, ${\cal F}\left(w\right)$ is not
known. This is important since the differential
decay  rate actually vanishes at zero recoil  (since there
is no phase space there). Experimentally, the
decay rate is measured as a function of
$w$ and extrapolated to
zero recoil using the form  of a Taylor expansion for the form factor
\begin{eqnarray}
{\cal F}\left(w\right) & = & {\cal F}\left(1\right)\left[1-\hat\rho^2\left(w-
1\right)+\hat c\left(w-1\right)^2+...\right]
\end{eqnarray}
Experiments are not yet sensitive to the quadratic term in the expansion and typically
quote values of $|V_{cb}|$ from a linear fit.  In the fits, the 
intercept and slope are highly correlated and a small correction to $|V_{cb}|$ of
order $1 \times 10^{-3}$ 
must be applied to compensate for the lack of curvature in the fit.
\end{enumerate}

\begin{figure}
\centerline{\psfig{figure=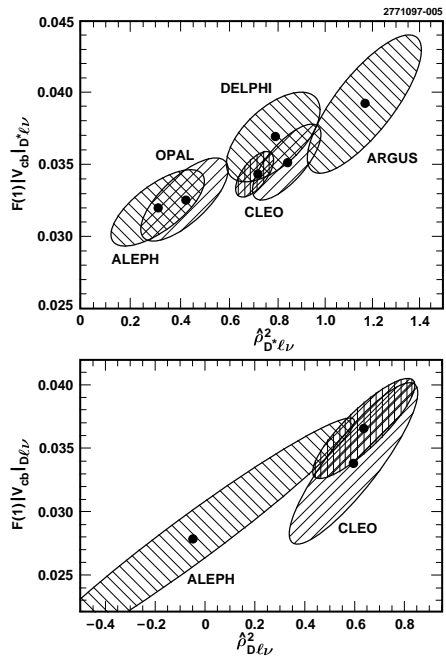,height=3.4in}}
\caption{One standard deviation error ellipses for  ${\cal F}\left(1\right)|V_{cb}|$
versus the form factor slope $\hat{\rho}^2$.  The black cross--hatched ellipse is
the world average.  The upper plot is for the decay $\bar B \to D^* \ell^-\bar{\nu}$
and the lower plot is for the decay $\bar B \to D \ell^-\bar{\nu}$.
\label{fig:vcb}}
\end{figure}

I follow the procedure outlined by Gibbons at Warsaw ~\cite{lkg}
to
extract a world average value for $|V_{cb}|$ from exclusive semileptonic
$B$ decays~\cite{lkg-new}. The experimental 
values for the intercept and the slope in these fits
~\cite{ds-exc-vcb}, summarized in
table ~\ref{tab:vcbrhodata} and  figure 
~\ref{fig:vcb}, are combined with careful attention to
the correlated errors yielding
the values in table ~\ref{tab:vcbrho}.

\begin{table}[t]
\caption{The experimental values of ${\cal F}(1)|V_{cb}|$ and the slope, $\hat{\rho}^2$,
 of the form factor as extracted from exclusive semileptonic
$B$ decays \protect \cite{ds-exc-vcb}.  $\tau_{B^0} = 
(1.55 \pm 0.04)\times 10^{-12}$ and
$\tau_-/\tau_0 = 1.06 \pm 0.04$ \protect \cite{schneider} \protect
are used in converting the branching 
fractions into  rates.
\label{tab:vcbrhodata}}
\vspace{0.4cm}
\begin{center}
\begin{tabular}{|c|c|c|c|}
\hline
 & & &  \\
 Mode & Experiment & ${\cal F}(1)|V_{cb}|$ & $\hat{\rho}^2$ \\
 &   & & \\ 
 \hline
 &  & & \\ 
$\bar B \rightarrow D^* \ell^-\bar{\nu}$ & ALEPH 
& $ 0.03205 \pm 0.0018 \pm 0.0019$ & $0.31 \pm 0.17 \pm 0.08$ \\
 & DELPHI & $0.0369 \pm 0.00209 \pm 0.00217$ & $0.782 \pm 0.187 \pm 0.036 $ \\
 & OPAL & $0.0326 \pm 0.0017 \pm 0.0022$ & $ 0.42 \pm 0.17 \pm 0.052 $ \\
 & CLEO & $0.03516 \pm 0.0019 \pm 0.00184$ & $ 0.84 \pm 0.12 \pm 0.08$ \\
 & ARGUS & $0.0392 \pm 0.0039 \pm 0.0028$ & $ 1.17 \pm 0.24 \pm 0.06$ \\
$\bar B \rightarrow D \ell^-\bar{\nu}$ & ALEPH  & 
$0.0282 \pm 0.0068 \pm 0.0065$ & $-0.05 \pm 0.53 \pm 0.38$ \\
& CLEO &$ 0.0342 \pm 0.0044 \pm 0.0049 $&$0.61 \pm 0.21 \pm 0.13$ \\
 & & & \\
\hline
\end{tabular}
\end{center}
\end{table}

\begin{table}[t]
\caption{The world average
values of $|V_{cb}|$ and the slope, $\hat{\rho}^2$,
 of the form factor ${\cal F}(1)$ as extracted from exclusive semileptonic
$B$ decays.  $\tau_{B^0} = (1.55 \pm 0.04) \times 10^{-12}$ and
$\tau_-/\tau_0 = 1.06 \pm 0.04$ \protect \cite{schneider} \protect
are used in converting the branching 
fractions into  rates.
\label{tab:vcbrho}}
\vspace{0.4cm}
\begin{center}
\begin{tabular}{|c|c|c|}
\hline
 & &  \\
 Mode & $|V_{cb}|$ & $\hat{\rho}^2$ \\
 &   &  \\ 
 \hline
 &  & \\ 
$\bar B \rightarrow D^* \ell^-\bar{\nu}$ & $ 0.0387 \pm 0.0031$ & $ 0.71 \pm 0.11$ \\
$\bar B \rightarrow D \ell^-\bar{\nu}$ & $ 0.0394 
\pm 0.0050$ & $ 0.66 \pm 0.19$ \\
 & & \\
\hline
\end{tabular}
\end{center}
\end{table}

Inclusive semileptonic decays are also used to extract $|V_{cb}|$. The dominant 
uncertainties in this procedure have traditionally been theoretical with 
typically ten percent errors assigned to the 
calculation of the rate,
 resulting in 
five percent errors on $|V_{cb}|$. 

A series of theoretical papers have taken the formalism of HQET in 
combination with the techniques of the operator
product expansion and shown that the inclusive $b\to 
c\ell^-\bar{\nu}$ rate is the leading term in a well-defined double expansion in 
$\alpha_s$ and $\Lambda_{\rm QCD}/M_{\rm Q}$, where the coefficients 
of the expansion involve matrix elements that reflect non perturbative effects
~\cite{inclvcb}. 
By experimentally measuring the moments of either the inclusive lepton 
spectrum or the hadronic mass spectrum squared in $b\to c\ell^-
\bar{\nu}$ decays, 
these unknown matrix elements can be experimentally determined and then 
used, in combination with the measured inclusive semileptonic rate, to 
extract $|V_{cb}|$.
I believe this offers real hope in improving the theoretical uncertainty on the
extraction of $|V_{cb}|$ from the inclusive semileptonic rate, and experimental
attempts to apply this formalism to data are in progress.

\begin{figure}
\centerline{\psfig{figure=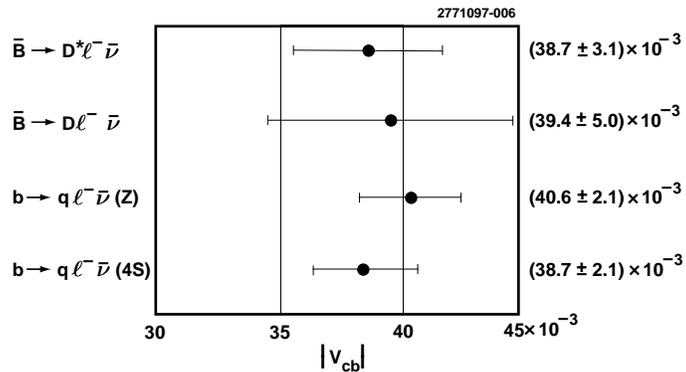,height=2.5in}}
\caption{The world average values of $|V_{cb}|$ from a variety of techniques.
The exclusive averages are described in this paper.  The inclusive averages
have not changed in the past year \protect \cite{lkg} \protect.
\label{fig:vcbsum}}
\end{figure}

Figure ~\ref{fig:vcbsum} gives the world average values of $|V_{cb}|$ extracted using
both inclusive and exclusive techniques.  The excellent agreement between 
a wide variety of methods for 
extracting $|V_{cb}|$ is quite encouraging; however, given the broad range of 
estimates for the theoretical uncertainties and the primitive status of many of 
the experimental checks of theoretical inputs, I advise against averaging the 
results.

\subsubsection{Checks of HQET}

It is extremely important that we experimentally check the predictions of HQET. 
Several approaches to this have been used so far, although they are not yet 
precise enough to be considered true tests.

HQET predicts simple relations between the three form factors, $A_1, A_2$ and
$V$ that are 
needed to describe the hadronic current in 
$\bar B\to D^*\ell^-\bar{\nu}$ decays.
These are usually expressed in terms of the form
factor ratios, $R_1(w)$ and $R_2(w)$.   By 
studying the full correlated angular distribution for these
decays, CLEO has shown consistency 
between experimental data ~\cite{cleoff} and the predictions of heavy quark 
symmetry ~\cite{neubert,candw,isgw2}
as summarized in table ~\ref{tab:ff}.
(Note: the interested reader can find the full expressions for the decay 
rate in terms of the form factors in a variety of excellent references. 
I particularly recommend Richman ~\cite{richman} and 
Richman and Burchat ~\cite {bandr}.)

\begin{table}[t]
\caption{Form factor ratios in $\bar B\rightarrow D^*\ell^-\bar{\nu}$ decays as 
measured by CLEO.
\label{tab:ff}}
\vspace{0.4cm}
\begin{center}
\begin{tabular}{|c|c|c|}
\hline
 & & \\
  & $R_1(w=1)$ & $R_2(w=1)$ \\
 &  &  \\ 
 \hline
 &  & \\ 
CLEO II~\cite{cleoff}& $1.24 \pm 0.26 \pm 0.12$ & $0.72 \pm 0.18 \pm 0.07$ \\
Neubert ~\cite{neubert} & $1.3 \pm 0.1$  & $ 0.8\pm 0.2$ \\
Close and Wambach ~\cite{candw} & 1.15 &  0.91 \\
ISGW2 ~\cite{isgw2}  & 1.27 & 1.01 \\
 & & \\
\hline
\end{tabular}
\end{center}
\end{table}

A second approach to  checking HQET is to compare the slope of the 
form factor distribution, ${\cal F}(w)$,  extracted from 
$\bar B \to D\ell^-\bar{\nu}$ and 
$\bar B \to D^*\ell^-\bar{\nu}$ decays.
These slopes, which are given in table ~\ref{tab:vcbrho},
 should agree in the heavy quark limit, although
finite mass corrections will be different for the two modes. The 
overall agreement between the slopes
of ${\cal F}(w)$ for $\bar B \to D\ell^-\bar{\nu}$ and $\bar B 
\to D^*\ell^-\bar{\nu}$ is encouraging
although the errors are large.

Yet another check is to compare $\hat{\rho}^2$, the slope of 
${\cal F}(w)$ extracted from the
$d\Gamma/dw$ distribution and given in table ~\ref{tab:vcbrho},
with the form factor slope, $\rho_{A_1}^2 = 0.91 \pm 0.016$
that is extracted from  fits to the full differential
distribution ~\cite{cleoff}
for $\bar B \to D^* \ell^-\bar{\nu}$ decay. The slope of ${\cal F}(w)$
has a complicated 
dependence on all three hadronic form factors involved in the decay:
$\hat{\rho}^2 = \rho_{A_1}^2 - f(R_1,R_2)$. Model 
dependent calculations ~\cite{neubertrhohat} suggest the relation 
$\hat\rho^2 \approx \rho^2_{A_1}- 0.2$.   The data
are consistent with this prediction within large errors.

\begin{figure}
\centerline{\psfig{figure=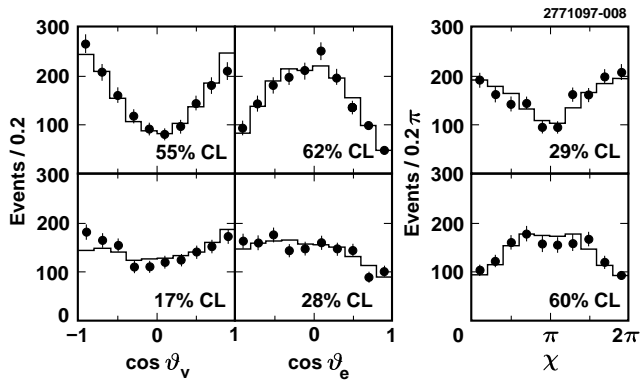,height=2.0in}}
\caption{The kinematic distributions in $D \to \bar{K^*} \ell^+\nu$
decay that are fit in the E791 analysis.  The top row of plots
is for the $q^2$ range: $q^2/q^2_{max} \leq 0.5$ and the bottom
row of plots is for $q^2/q^2_{max} > 0.5$.
The points are data and the histogram is the result of the fit.
\label{fig:e791}}
\end{figure}

Finally, we can attempt to use the same machinery developed for 
$\bar B\to  D^{(*)}\ell^-\bar{\nu}$ decays on $D\to 
\bar K^{(*)}\ell^+\nu$ decays. 
The CKM matrix element for these charm semileptonic decays is known so that the form
factor can be measured directly.
Very clean signals  are obtained in the charm hadroproduction fixed
target experiment E791 taking  full advantage of their 23 planes of high
resolution silicon strips to select  separated charm vertices ~\cite{e791}.
They reconstruct 3000 $D\to \bar K^*\ell^+ \nu$ decays with very little background
and then extract the form factors by doing a 4 dimensional fit to the
kinematic variables describing the decay.
Figure ~\ref{fig:e791} shows the kinematic distributions that are fit in the
E791 analysis. Again, the interested reader is referred
to several excellent
papers ~\cite{richman,bandr} for the  full expressions for the decay 
rate in terms of the form factors. 
While the $s$ quark does not probably classify as heavy and so $\Lambda/M_Q$ 
corrections will be large,  the new results from E791 show quite good agreement
with recent lattice calculations of the form
factors as summarized in table ~\ref{tab:e791ffsum}.

\begin{table}[t]
\caption{Form factor ratios in $D\rightarrow \bar{K^*}\ell^+ \nu$ decays, as 
measured by E791, compared with various calculations.
\label{tab:e791ffsum}}
\vspace{0.4cm}
\begin{center}
\begin{tabular}{|c|c|c|c|}
\hline
 & & & \\
  & $A_1(0)$ & $A_2(0)$ & $V(0)$ \\
 &  &  & \\ 
 \hline
 &  &  & \\ 
E791 ~\cite{e791} & $0.58 \pm 0.03$ & $0.41 \pm 0.06$ & $1.06 \pm 
0.09$ \\
APE ~\cite{ape} & $0.67 \pm 0.11$ & $0.49 \pm 0.34$ & $ 1.08 \pm 0.22$ \\
Wupp ~\cite{wupp} & $0.64 \pm 0.06$ & $0.61 \pm 0.41$ & $1.17 \pm 0.38$ \\
UKQCD ~\cite{ukqcd} & $0.70 ^{+0.07}_{-0.10}$ & $0.66^{+0.10}_{-0.15}$ & 
$1.01^{+0.30}_{-0.13}$ \\
ELC ~\cite{elc} & $0.64 \pm 0.16$ & $ 0.41\pm 0.28$ & $0.86 \pm 0.24$ \\
\hline
 & & & \\
  & $A_1(q^2_{max})$ & $A_2((q^2_{max})$ & $V((q^2_{max})$ \\
 &  &  & \\ 
 \hline
 &  &  & \\ 
E791 ~\cite{e791} & $0.68 \pm 0.04$ & $0.48 \pm 0.08$ & $1.35\pm0.12$ \\
ISGW2 ~\cite{isgw2} & 0.70 & 0.94 & 1.52 \\
 & & & \\
\hline
\end{tabular}
\end{center}
\end{table}

I conclude that while we cannot claim yet that HQET has been 
experimentally verified by testing the $1/M_Q$ corrections
to it, the exclusive semileptonic decays of bottom to charm are reasonably
well 
understood experimentally and theoretically at this point.

\subsection{Semileptonic Charmless $B$  Decays and $|V_{ub}|$}

\begin{figure}
\centerline{\psfig{figure=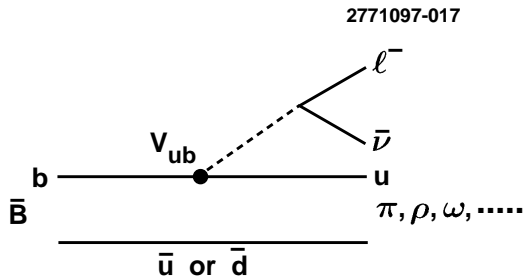,height=1.5in}}
\caption{Diagram illustrating the semileptonic decays of a $B$ meson to charmless
final states.
\label{fig:btou}}
\end{figure}

Semileptonic $B$ decays to charmless final states are used to measure 
$|V_{ub}|$. The basic process is illustrated
in figure ~\ref{fig:btou} and is the same as for semileptonic $b\to c\ell^-\bar{\nu}$ decays 
except now the $b$ quark turns into a $u$ quark with coupling strength 
$|V_{ub}|$.

\subsubsection{Exclusive and Inclusive Charmless $B$ Semileptonic Branching Ratios}

To date, inclusive searches for $b\to u \ell^-\bar{\nu}$ decays have been done at the 
$\Upsilon\left(4S\right)$ where $B$ mesons are produced at rest
~\cite{upsilon4sbtou}. These
analyses have 
looked at the endpoint of the
single lepton spectrum for leptons from $B$ decay that are 
kinematically incompatible with coming from the decay of a  $B$ meson
to a charm meson. 
The $b$ quark will 
preferentially turn into a $c$ quark when it weakly decays, but $c$ quarks
 are 
heavy. The lightest mass particle containing a charm quark is a $D$ meson 
which has a mass of 1.8 GeV. In charmless decays, the final state 
hadronic mass can be as light as a pion mass. The difference in the final state 
hadronic mass is reflected in the momentum of leptons from the decaying 
$B$;  for a $B$ meson at rest,
 the endpoint spectrum for leptons in charmless semileptonic 
decays 
extends about 300 MeV past the endpoint spectrum for semileptonic decays
to charm.
From the lepton excess in the endpoint region, one uses a model to extrapolate the 
full spectrum and extract:
\begin{eqnarray}
|V_{ub}| & = & (3.1 \pm 0.8) \times 10^{-3}
\end{eqnarray}
The model uncertainty dominates the error ~\cite{patterson}.  

\begin{figure}
\centerline{\psfig{figure=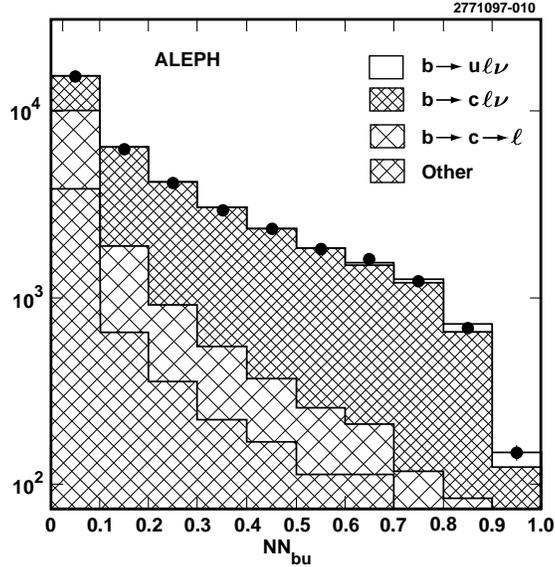,height=3.0in}}
\caption{The neural net output indicating the presence of 
$b\rightarrow u \ell^-\bar{\nu}$ events in the ALEPH data.
\label{fig:alephbtou}}
\end{figure}

A new inclusive analysis was presented by ALEPH which is the first evidence 
for semileptonic $b\to u$ transitions in {\it b}-hadrons produced at LEP
~\cite{alephbtou}. ALEPH
inclusively
reconstructs the hadronic system  accompanying the lepton in the semileptonic
$B$ decay and builds a set of  kinematic variables to discriminate between
$X_u\ell^-\bar{\nu}$ and 
$X_c\ell^-\bar{\nu}$ transitions by taking advantage of the different shape 
properties of these final states. The analysis is very aggressive, using a neural 
network technique to extract the inclusive $B\to X_u\ell^-\bar{\nu}$ branching 
ratio. An advantage of the technique is that it integrates over the entire 
lepton and hadron spectra for these decays, a technique not possible in
$\Upsilon\left(4S\right)$ measurements, potentially reducing the model
dependence of  the result. A disadvantage is that the large background from
$b \rightarrow c$ semileptonic decays needs to be understood at the one percent level for a
meaningful extraction of $|V_{ub}|$~\cite{patterson} since the
boost of the $B$ mesons at LEP is sufficient  that even in the endpoint
of the lepton spectrum, $b \to u \ell^-\bar{\nu}$ and $b \to c \ell^-\bar{\nu}$
events cannot be cleanly separated. Systematic errors on
the analysis are still
being evaluated; however, figure ~\ref{fig:alephbtou} shows the neural net
output from the ALEPH analysis indicating the presence of a $b \rightarrow u \ell^-\bar{\nu}$
signal.

$|V_{ub}|$ has also been extracted from the exclusive
decays $\bar{B}\to\pi\ell^-\bar{\nu}$ and $\bar{B}\to\rho\ell^-\bar{\nu}$ at CLEO that were  first reported
last year ~\cite{cleopilnu}.  From the exclusive $\bar{B}\to\pi\ell^-\bar{\nu}$ and
 $\bar{B}\to\rho\ell^-\bar{\nu}$ rates, the value is extracted: 
\begin{eqnarray}
|V_{ub}| & = & (3.3 \pm 0.2^{+0.3}_{-0.4} \pm 0.7) \times 10^{-3}
\end{eqnarray}
where the errors are statistical, experimental systematic, and
theoretical model dependence respectively.

All $\left|V_{ub}\right|$ extractions rely heavily on models. The 
consistency of the results is very encouraging, but the large error bars are still 
dominated by theoretical uncertainties with poorly understood errors.

\subsubsection{HQET and Charmless $B$ Decays}

Heavy quark symmetry is not obviously helpful in the analyses of semileptonic
$b\to u$ 
decays since the light quark in the initial state is most certainly aware that the 
quark it is bound to in the final state is no longer heavy. The overlap integral
is analogous to  the overlap between the electron wave function in hydrogen and
in  positronium. However, HQET  can be used to relate
the  form factors for $c\to d$ semileptonic decays to those for $b\to u$ 
semileptonic decays at the same $q^2$.  In those two cases, the initial 
state and final state wave functions will be very similar.  
Since $|V_{cd}|$ is known from neutrino production of charm, form factors 
can be measured and models tested in charm decays and then extended to the 
$B$ system to reduce theoretical errors on the $|V_{ub}|$ extraction ~\cite{hqet}.

New results in the charm sector from the
fixed target experiments E687 and E791, as well as new
CLEO results, have been encouraging. New 
measurements of ${\cal B}(D^0\to\pi^-\ell^+\nu)$ relative to 
${\cal B}(D^0\to K^-\ell^+\nu)$ and ${\cal B}(D^+\to\pi^0\ell^+\nu)$ relative to 
${\cal B}(D^+\to\bar{K^0}\ell^+\nu)$, as listed in table ~\ref{tab:psff},
can be used to extract the ratio of hadronic form factors for 
the pseudoscalar to pseudoscalar decays.

\begin{table}[t]
\caption{Ratios of branching ratios for pseudoscalar to
pseudoscalar semileptonic charm decays.
\label{tab:psff}}
\vspace{0.4cm}
\begin{center}
\begin{tabular}{|c|c|c|}
\hline
 & & \\
Experiment &${\cal B}(D^0\to\pi^-\ell^+\nu)/{\cal B}(D^0\to K^-\ell^+\nu)$
 &${\cal B}(D^+\to\pi^0\ell^+\nu)/{\cal B}(D^+\to \bar{K^0}\ell^+\nu)$  \\
 &  &   \\ 
 \hline
 &  &   \\ 
Mark III ~\cite{mkiii} & $0.11 ^{+0.07}_{-0.04}\pm 0.02$ & \\
CLEO ~\cite{cleopsps} & $0.103 \pm 0.039 \pm 0.013$ & $0.046 \pm 0.014 \pm 0.017$\\
E687 ~\cite{e687} & $0.101 \pm 0.020 \pm 0.003$ & \\
 &  & \\
\hline
\end{tabular}
\end{center}
\end{table}

\begin{figure}
\centerline{\psfig{figure=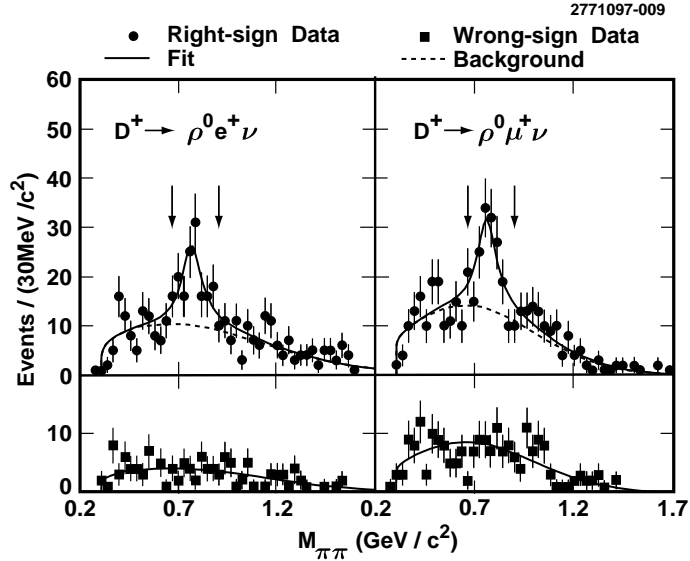,height=3.0in}}
\caption{E791 signals for $D^+\to\rho^0\ell^+\nu $.
\label{fig:e791cs}}
\end{figure}

For the pseudoscalar to vector semileptonic decays, 
E791 reports a new measurement of the branching ratio 
${\cal B}\left(D^+\to\rho^0\ell^+\nu\right)/ $ 
${\cal B}\left(D^+\to
\bar K^{*0}\ell^+\nu\right)$ which is mainly sensitive to the $A_1$ form
factor ~\cite{e791dtorho}. Figure ~\ref{fig:e791cs} shows
the very clean signals that are obtained 
with their
23 planes of high resolution silicon strips.
The E791 and E687  results are beginning to
discriminate among models, as summarized in table
~\ref{tab:dtorhoff}, that are also used to predict form  factors for $b\to
u$ semileptonic decays to extract $|V_{ub}|$.  However, 
there is much more progress to be made and much larger data sets are needed 
before we will have a precision measurement of $|V_{ub}|$.

\begin{table}[t]
\caption{Ratios of rates for pseudoscalar to
vector semileptonic charm decays. Most theoretical 
results are calculated for $D^0$ decays so following
experimental references, 
the theoretical results for
$D^0$ decay are compared to experimental results for $D^+$ decay
using the relations: $\Gamma(D^+\to \bar K^{*0}\ell^+\nu)=
\Gamma(D^0\to K^{*-}\ell^+\nu)$ and $\Gamma\left(D^+\to\rho^0\ell^+\nu\right)
=\frac{1}{2}\times\left(D^0\to\rho^-\ell^+\nu\right)$.
\label{tab:dtorhoff}}
\vspace{0.4cm}
\begin{center}
\begin{tabular}{|c|c|}
\hline
 & \\
Experiment & ${\cal B}\left(D^+\to\rho^0\ell^+\nu\right)/ $ 
${\cal B}\left(D^+\to
\bar K^{*0}\ell^+\nu\right)$ \\
   &   \\ 
 \hline
   &   \\ 
E653 ~\cite{e653}  & $0.044 ^{+0.031}_{-0.025}\pm 0.014$ \\
E791 ~\cite{e791dtorho} & $0.047 \pm 0.013$ \\
E687 ~\cite{e687rho} & $0.073 \pm 0.019  \pm 0.013 $ \\ 
  & \\
\hline 
 & \\
Theory & ${\cal B}\left(D^+\to\rho^0\ell^+\nu\right)/ $ 
${\cal B}\left(D^+\to
\bar K^{*0}\ell^+\nu\right)$ \\
 & \\
\hline
  & \\
ISGW2 ~\cite{isgw2} & 0.022 \\
Jaus ~\cite{jaus} & 0.030 \\
BSW ~\cite{bsw} & 0.037 \\
ELC ~\cite{elc} & $0.047 \pm 0.032$ \\
APE ~\cite{ape} & $0.043 \pm 0.018 $\\
UKQED ~\cite{ukqcd} & $0.036 ^{+0.010}_{-0.013}$ \\
LMMS ~\cite{lmms} & $0.040 \pm 0.011$ \\
Casalbuoni ~\cite{casalbuoni} & 0.06 \\
 & \\
\hline
\end{tabular}
\end{center}
\end{table}

\subsection{Puzzles in Semileptonic $B$ Decay}

There are 2 puzzles in semileptonic $B$ decay. Neither is 
significant enough to rate the label of discrepancy and new results this year 
did not resolve either puzzle.

\subsubsection{Puzzle 1}
The first puzzle is that there has persistently been a difference
in the
semileptonic 
$b$ branching ratios measured at the $\Upsilon\left(4S\right)$ and the $Z$ 
with the value from LEP being higher as is evident in 
table ~\ref{tab:slbr}.  The ratio of branching ratios
is 
\begin{eqnarray}
R = \frac{{\cal B}^{4S}_{SL}}{{\cal B}^{Z}_{SL}} & = & 0.930 \pm 
0.046
\end{eqnarray}
Given the short $\Lambda_B$ hadron lifetimes
measured at LEP, and if we assume that the semileptonic rates
are similar for all species of $B$ hadrons, 
we would actually expect the LEP value for the
semileptonic $b$ branching ratio to be lower than that measured at the 
$\Upsilon\left(4S\right)$, making the actual 
disagreement more severe. At this
level of disagreement, about 2 standard
deviations, one can only take a ``wait and see'' attitude. It is not clear if this
is a  problem or not.

\subsubsection{Puzzle 2}
The second puzzle is that historically, persisting to the present day, theoretical 
predictions of the semileptonic $B$ branching ratio have been 
 significantly
larger than the experimentally
measured values.
The average
experimental semileptonic branching ratios are given in table ~\ref{tab:slbr}.
Traditionally, the lower limit on the theoretical value for $B_{SL}$ has
been $12.5$ percent ~\cite{bigi}, which disagrees with the experimental values by many
standard deviations.

If we believe the experimental numbers, and I do---they
have been stable for a number of years and have been measured with a
sufficient variety of techniques that it is hard to suspect serious
experimental problems---then we must examine the ingredients of the theoretical
calculations to see where the problem might lie.

We can write

\begin{eqnarray}
{\cal B}_{SL}& = &{\Gamma_{\rm semi-leptonic} \over \Gamma_
{\rm semi-leptonic} +
\Gamma_{\rm hadronic} + \Gamma_{\rm leptonic}}.
\end{eqnarray}

\noindent $\Gamma_{\rm leptonic}$ is very small,
and I have argued that we understand the semileptonic rate so the likely
culprit is a flaw in our understanding of $\Gamma_{\rm hadronic}$ for $B$ meson
decays.

If we break hadronic rate into its constituent pieces, we have
$
\Gamma_{\rm hadronic}\left( b\right)  = \Gamma\left( b\to c\bar ud\right) +
\Gamma
\left(b\to c\bar cs\right) + \Gamma\left( b\to sg\right)
$ ($g$ here refers to $glue$) 
and a variety of calculations have examined these component rates in
detail. Note that to reduce the expectation for the semileptonic branching
ratio from 12.5 \% to the experimental value of 10-11\%  requires a 20\%
enhancement of $\Gamma_{\rm hadronic}$ so we presumably are not looking for
something terribly subtle!

Theoretical solutions to the problem fall into four categories.

\begin{enumerate}

\item There could be an enhancement of $\Gamma\left(b\to c\bar ud\right)$ due to
non-perturbative effects. The problem with this class of solutions is that it
predicts $\tau_{B^+}/\tau_{B^0}\sim 0.8$,~\cite{klausratio}
which is not a
comfortable fit to the experimental lifetime ratio of $1.06\pm0.04$~\cite{schneider}.

\item There could be an enhancement of $\Gamma\left(b\to c\bar cs\right)$ due 
to large QCD
corrections. The problem with this solution is that it affects another
experimental quantity, $n_c$, which is the average number of charm or
anticharm quarks per $b$ decay. Theoretical models that significantly 
increase $\Gamma\left(b\to
c\bar cs\right)$ to bring the semileptonic branching ratio 
into line with experiment
find $n_c$ in the range of 1.2 to 1.3~\cite{morenc,nc}.

\end{enumerate}

\begin{figure}
\centerline{\psfig{figure=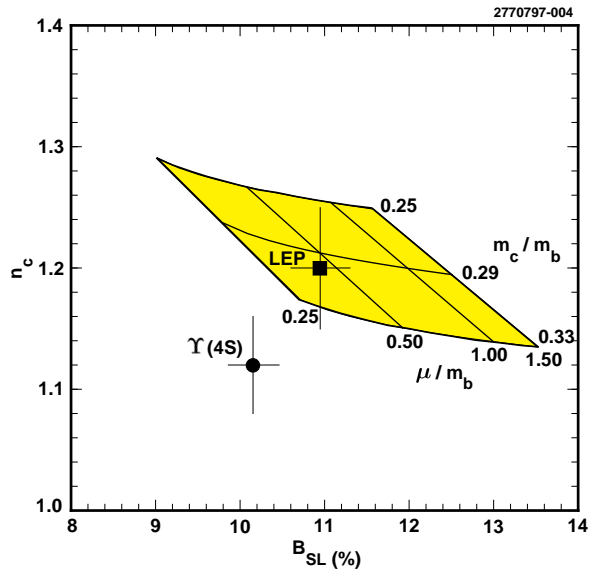,height=3.0in}}
\caption{The number of charm quarks per $b$ decay vs. the semileptonic branching
ratio.  The theoretical curves are from Neubert
and Sachrajda \protect \cite{nc} \protect.
The experimental
values are the averages compiled for this review.
\label{fig:nc}}
\end{figure}

Experimentally, $n_c$ looks to be lower than the theoretical values
although when both theoretical errors and experimental errors
are included, this solution looks appealing as summarized in 
figure ~\ref{fig:nc}.  However,
it is important to note that the measurements that make up $n_c$ are in
disagreement with each other at roughly the 2 sigma level in how many $D$
mesons there are per $b$ decay, so that it is perhaps premature to 
talk about either agreeing or disagreeing with theory at this point.

Table ~\ref{tab:nc} lists the branching ratios from LEP
and  from CLEO  that are summed to extract $n_c$.
We see that there are more $D_s$'s and charm-baryons per $b$ quark
at LEP as to be expected, since at LEP the $b$ quark forms a $B_s$ or
$\Lambda_b$ a fraction $f_{b\rightarrow B_s}$ or
$f_{b\rightarrow \Lambda_B}$ of the time.
At CLEO, only $B_u$ and $B_d$ can be produced.  What is
problematic is that one would expect to see fewer $D$ mesons per $b$ 
decay at LEP (for just the same reason...since $b$ quarks are forming $B_s$ or
$\Lambda_b$, then only a fraction of the
$b$ quarks, $f_{b\rightarrow B}$, are 
making $B_u$ and $B_d$ which decay dominantly to $D$ mesons) and the measured
rates for $b \to DX$ at CLEO and LEP don't show enough difference.  The
numbers are too close, if we assume the traditional values
$f_{b\rightarrow B} = 0.378 \pm 0.022$, $f_{b\rightarrow B_s} = 0.112 \pm 0.019$,
and $f_{b\rightarrow \Lambda_B} = 0.132 \pm 0.041$ ~\cite{pdg}!

\begin{table}[t]
\caption{Summary of the branching ratios used in the calculation
of $n_c$.  The $\Upsilon(4S)$ branching ratios are from CLEO 
\protect \cite{cleobtod,browandh,browhf} \protect.  The LEP 
branching ratios are averaged over the four LEP experiments with careful
attention to correlated systematic errors \protect \cite{lepbtod} \protect, and all branching ratios
are scaled to the normalizing branching ratios listed in table 
\protect \ref{tab:dbr} \protect.
\label{tab:nc}}
\vspace{0.4cm}
\begin{center}
\begin{tabular}{|c|c|c|}
\hline
 & &  \\
  & $\Upsilon(4S)$  &  LEP \\
 &    & \\ 
 \hline
 &    & \\ 
${\cal B}(b \rightarrow D^0X)$  & $ 0.642 \pm 0.030$  & $0.576 \pm 0.026 $\\
${\cal B}(b \rightarrow D^+X)$  & $ 0.246 \pm 0.021$  & $0.224 \pm 0.019$ \\
${\cal B}(b \rightarrow D_s^+X)$  & $ 0.118 \pm 0.031$  & $0.191 \pm 0.050$ \\
${\cal B}(b \rightarrow \Lambda_C^+X)$  & $ 0.039 \pm 0.020$  & $0.114 \pm 0.020 $\\
${\cal B}(b \rightarrow \Xi_cX)$  & $ 0.020 \pm 0.010$  & $0.063 \pm 0.021$ \\
${\cal B}(b \rightarrow (c\bar{c})X)$  & $ 0.054 \pm 0.007$  & $0.034 \pm 0.012$ \\
  & & \\
\hline
 & &  \\
$n_c = \Sigma{\cal B}_i$  & $1.119 \pm 0.053$ & $1.202 \pm 0.067$ \\
 &    & \\ 
 \hline
\end{tabular}
\end{center}
\end{table}

I estimate this discrepancy to be at the 1.8 $\sigma$ level
when I carefully include all errors (correlated and
uncorrelated) and scale the experimental results
to common D branching ratios as listed in table ~\ref{tab:dbr}.

This discrepancy is just at the
level that it might make you uncomfortable and it is certainly
in need of resolution before the experimentalists can claim any serious
conflicts with theory.

\begin{table}[t]
\caption{Summary of the normalizing $D$ branching ratios used in the calculation
of $n_c$.
\label{tab:dbr}}
\vspace{0.4cm}
\begin{center}
\begin{tabular}{|c|c|}
\hline
 &  \\
${\cal B}(D^0\rightarrow K^-\pi^+)$ ~\cite{richman} & $0.0388 \pm 0.0010$ \\
${\cal B}(D^+\rightarrow K^-\pi^+\pi^+)$ ~\cite{richman} & $0.088 \pm 0.006$ \\
${\cal B}(D_s^+\rightarrow\phi\pi^+)$ ~\cite{pdg} & $0.036\pm 0.009$ \\
  & \\ 
 \hline
\end{tabular}
\end{center}
\end{table}

One implication, if this discrepancy holds up
under further investigation, is that perhaps there is an  error in our understanding
of how often the b quark makes a $B$ meson at LEP!
Turning the comparison of these branching ratios into a measurement of
$f_{b\rightarrow B}$ gives $f_{b\rightarrow B} = 0.45 \pm 0.03$
to be compared with $0.378\pm 0.022$ which is traditionally used.
It is amusing to note that new measurements of $f_{b\rightarrow B}$
reported at this conference are helpful in resolving this problem ~\cite{schneider}.

\begin{enumerate}
\setcounter{enumi}{2}
\item In a variation on the previous solution to the problem,
it has been suggested that a
sizable fraction of $b \to c\bar c s$ transitions
could appear as $b \to {\rm no~open~charm}$~\cite{hitoshi}.
The hypothesis is that a
large component of low mass $c\bar c$ pairs are
seen as light hadrons and not as open charm.  $n_c$
would not be increased by this mechanism, and this possibility
 is currently under experimental
investigation.

\item The most intriguing possibility for
solving the semileptonic branching ratio experimental
shortfall is an enhancement of $\Gamma\left(b\to
sg\right)$ from some unexpected source of new physics.
\end{enumerate}

Theorists also point out a fifth solution.
\begin{enumerate}
\setcounter{enumi}{4}
\item There could be a systematic experimental problem. One place attention
has focussed is on the $D^0\to K^-\pi^+$ branching ratio ~\cite{isi}.

\end{enumerate}

\subsubsection{Experimental Attempts to Resolve Puzzles in Semileptonic $B$ Decays}
Given no compelling
theoretical resolution to  the
puzzles in the semileptonic rate, a variety of experiments in the last
year have attempted to address the issue. The most comprehensive new
studies are  done by 
CLEO ~\cite{cleo_brsol} and DELPHI ~\cite{delphi_brsol}.

 CLEO uses lepton-$D$ charge and angular
correlations to extract rates for $\bar{B}\to DX$, $\bar{B}\to\bar DX$, and $\bar{B} \to
DX\ell^-\bar{\nu}$. With these three measurements, three unknowns can be extracted:
\begin{enumerate}
\item the number of $D$'s produced at the upper vertex in $B$ decay:
${\cal B}(B \rightarrow D X) = 0.079 \pm 0.022$ 

\item ${\cal B}(b \rightarrow s g) = 0.002 \pm 0.040$ or $< 0.068$ at 90\% 
confidence level, and 

\item $ {\cal B}(D^0 \rightarrow K^-\pi^+) = 0.0369 \pm 0.0020$.
\end{enumerate}

The new CLEO measurement of ${\cal B}\left(B\to DX\right)$
 gives ${\cal B}\left(b\to c\bar cs\right) = {\cal B}
\left(b\to \left(c\bar c\right)s\right) + {\cal B}\left(b\to c D_s\right) +
{\cal B}\left(b\to {\rm charm-baryons}\right) + {\cal B}\left(b\to
DX\right) =
0.219\pm 0.036 $, in good
agreement with theoretical expectation.  The $D^0 \rightarrow K^-\pi^+$ branching
ratio extracted from the analysis is in good agreement with the world average
listed in table ~\ref{tab:dbr} and another new CLEO measurement of the
same branching ratio using a partial reconstruction technique ~\cite{cleoprdkpi},
and the new limit on $b \rightarrow s g$ neither supports nor rules out the 
possibility
of new physics.

The new DELPHI analysis fits to the b quark tagging probability distribution
to contributions from events with 0 charmed particles , 1 charmed
particle  or 2 charmed particles, where the different final states can
be distinguished since they have different numbers of secondary
vertices in the final
state.  DELPHI extracts the branching ratio to 0 charm (no open charm)
in the final state, and the branching ratio to 2 charm particles in the 
final state.  By subtracting  the  Standard Model expectation for the branching
ratio to no open
charm, they put a limit on new physics producing no open charm of $4.5\%$ at
the 95\% confidence level.

What do we conclude? I feel put in a miserable position as a reviewer since I can
conclude nothing. I am actually not totally convinced there is
a problem when all errors, both theoretical and experimental
are included.  Radical theoretical solutions seem experimentally
unacceptable.  Experimental
probes of the components that go into making up the rate see some experimental
problems, but no solutions to the
overall problem. There is no evidence for new physics and
no conclusive possibility of ruling it out. To nail this down 
and resolve this messy issue in a satisfactory manner will require more data.

\section{Heavy Quarks Beyond Tree Level}

So far, I have been concentrating on heavy meson decays at tree level. I now want
to switch to discussing heavy meson decays beyond tree level and rare
decays. I want to take a few minutes to stress the importance of these decays,
particularly in the $B$ sector.

Rare processes, for which the Standard Model expectation for the rate is small,
can be used to probe new physics. Those processes that are highly suppressed or
explicitly forbidden to first order in the Standard Model are especially
sensitive since often new
physics can compete favorably.
The other interest in $B$ decays beyond tree level is that they offer methods
for determining the phases of the elements of the CKM matrix, thereby probing the
Standard Model mechanism for CP violation.

I want to stress that in contrast to the results that dominated the
first part of this review, the new results on
hadronic rare $B$ decays are full of surprises. While we
can start to observe overall patterns, the patterns are largely unexplained.

\subsection{Rare Hadronic $B$ Decays}

This year has  seen a flood of new results on rare $B$ decays, especially
from the CLEO
experiment. Where previously there were upper limits, now in many cases there
are signals. Particularly for the hadronic rare $B$ decays, there have been
significant improvements in analysis techniques.   However, the dominant
improvement has been the addition of  more data so that the new
results presented this year are based on the full $3.3 \times 10^6 B\bar B$ of the CLEO II data
set.

\subsubsection{$B \to \pi\pi/\pi K/KK$}

\begin{figure}
\centerline{\psfig{figure=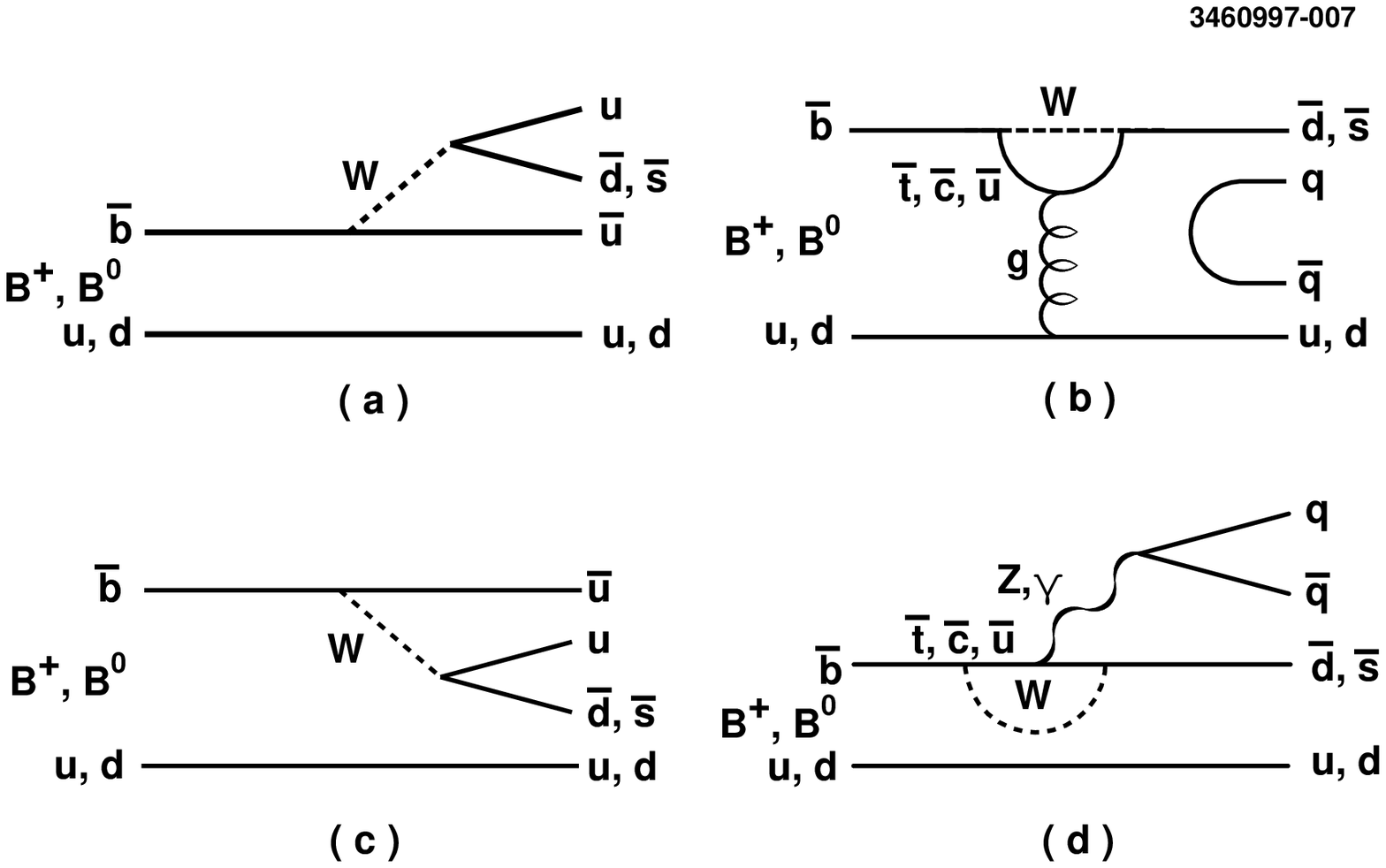,height=3.0in}}
\caption{Feynman diagrams illustrating the dominant decay
processes  that
contribute to the $B \to \pi\pi/K\pi$ decays: (a)external $W$-emission, (b) gluonic penguin, (c) internal
$W$-emission, and (d) external electroweak penguin.
\label{fig:kpipi}}
\end{figure}

Let me start by discussing the new CLEO results on the $B \to \pi\pi/K\pi/KK$ decay 
modes.
Many diagrams contribute to these decays as illustrated in figure ~\ref{fig:kpipi},
 but penguin and tree diagrams are
expected to dominate. 

\begin{figure}
\centerline{\psfig{figure=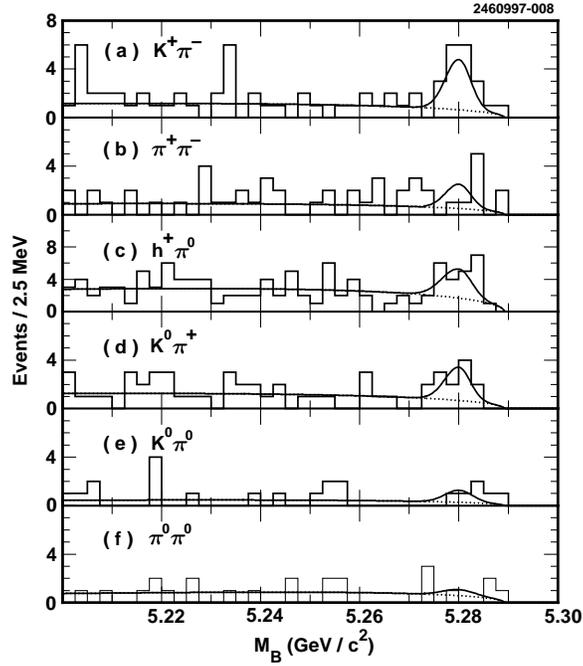,height=3.5in}}
\caption{Reconstructed $B$ mass plots for CLEO data
(a)$B^0 \to K^+\pi^-$, (b) $B^0 \to \pi^+\pi^-$,
(c) $B^+ \to h^+\pi^0$, (d) $B^+ \to \pi^+K^0$,
(e) $B^0 \to K^0\pi^0$,
and (f) $B^0 \to \pi^0\pi^0$.  The data are the histogram and the
scaled projection of the total likelihood fit (solid curve) and the continuum
background (dotted curve) are overlaid.  $h^+$ indicates 
 $K^+$ or $\pi^+$.
\label{fig:kpi}}
\end{figure}

The CLEO analysis looks in $\Upsilon(4S)$ decays 
for two stiff back-to-back particles. The
dominant backgrounds are $e^+e^-\to q\bar q$ from the continuum under the
$\Upsilon\left(4S\right)$ resonance with continuum charm accounting for about 25
percent of the background. Yields are extracted from a likelihood fit to the data
and branching ratios are calculated:
\begin{eqnarray}
{\cal B}(B^0 \rightarrow K^+\pi^-) & = & \left 
( 1.5 ^{+0.5 + 0.1}_{-0.4 - 0.1} \pm 0.1\right ) \times 10^{-5} \\
{\cal B}(B^0 \rightarrow \pi^+\pi^-) & < & 1.5 \times 10^{-5}  ~{\rm (90 \% c.l.)}\\
{\cal B}(B^0 \rightarrow K^+K^-) & < & 0.4 \times 10^{-5} ~{\rm (90 \% c.l.)}
\end{eqnarray}
Similar analyses yield significant signals in 
$B^+ \to \pi^+ K^0$ and in $B^+\to (\pi^+\pi^0
+K^+ \pi^0$), which is denoted $B \to h^+ \pi^0$. 
These results are summarized in  
table \ref{tab:kpisum}. Also indicated
are the dominant amplitudes thought to be contributing to each decay ~\cite{gronau}.
Figure ~\ref{fig:kpi} shows the projections in reconstructed $B$ mass for the signals.

\begin{table}[t]
\caption{Summary of the new CLEO results on the $B\to K\pi/\pi\pi/K\pi$ branching
ratios.  The second column indicates the dominant amplitudes for each decay using
the notation of 
Gronau {\it et al.} \protect \cite{gronau} \protect.  Upper limits are
quoted at 90 \% confidence level.  $h^\pm$ indicates $\pi^\pm$ or $K^\pm$.
\label{tab:kpisum}}
\vspace{0.4cm}
\begin{center}
\begin{tabular}{|c|c|c|}
\hline
 & & \\
Mode & Amplitude & ${\cal B} (10^{-5})$  \\
& & \\
\hline
& & \\
$B \to K^{\pm}\pi^{\pm}$ & $-(T' + P')$ & $1.5 ^
{+0.5 + 0.1}_{-0.4 - 0.1} \pm 0.1$    \\
$B \to K^\pm\pi^0 $  &    $-(T'+C'+P')/\sqrt 2$  & $<1.6$ \\
$B \to K^0\pi^\pm$  & $P'$  & $2.3 ^{+1.1 + 0.3}_{-1.0 -0.3} \pm 0.2$ \\
$B \to K^0\pi^0  $  & $-(C'-P')/\sqrt 2$ & $ <4.1$ \\
& & \\
\hline
& & \\
$B \to \pi^\pm \pi^\mp$  & $-(T + P)$   &  $<1.5$ \\
$B \to\pi^\pm \pi^0$  & $-(T+C)/\sqrt 2$ & $<2.0$ \\
$B \to\pi^0\pi^0$  & $-(C-P)/\sqrt 2$ & $<0.9$ \\
& & \\
\hline
& & \\
$B \to K^\pm K^\mp$  & $E$  &  $<0.4$ \\
$B \to K^\pm K^0$  & $P$  & $<2.1$ \\
$B \to  K^0 \bar{K}^0$  & $P$  & $<1.7$ \\
& & \\
\hline
& & \\
$B \to h^\pm \pi^0$  & & $1.6^{+0.6 + 0.3}_{-0.5 - 0.2} \pm 0.2$ \\
$B \to h^\pm K^0$ & & $2.4^{+1.1 + 0.2}_{-0.9 - 0.2} \pm 0.2$\\
 &  & \\ 
 \hline
\end{tabular}
\end{center}
\end{table}

The observation of $B \to K^0\pi^+$ is interesting because it directly measures the
strength of the gluonic penguin. The observed rate is about a factor of two
higher than expected. This is our first surprise. We now can look at
$B \to K^\pm\pi^\mp$ which both the tree and penguin can contribute to. The fact that
the branching ratio is smaller than $B\to \pi^+ K^0$ may indicate  interference and
Fleischer has used this to draw tentative conclusions about 
$\gamma$, the phase
of $V_{ub}$,
although statistical errors are
very large still~\cite{fleischer}. The branching ratio for
$B \to \pi^\pm\pi^\mp$ is small compared to expectation, and there is no explanation
why it is low. Perhaps it is a fluctuation since in comparison with
$B \to \pi^+\pi^0$ we expect twice the rate. The
$B \to KK$ modes are the one place where there are no surprises. 
We do not expect much, and we do not see much.

I want to briefly mention that both ALEPH and DELPHI have statistically
significant signals in the sum of $B^0 \to
(K^+\pi^-+\pi^+\pi^-)$ that are consistent
with and less precise than new CLEO results ~\cite{lepbpipi}.

\subsubsection{Other Rare Hadronic $B$ decays}

\begin{figure}
\centerline{\psfig{figure=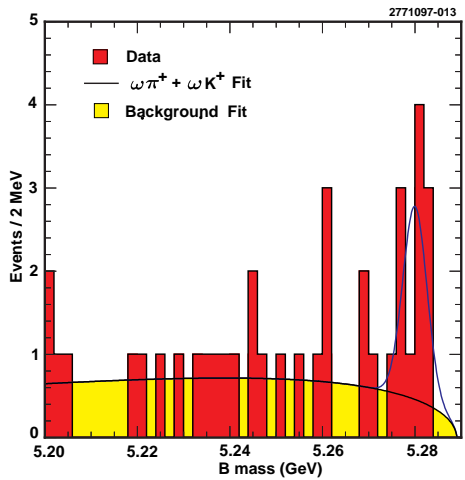,height=2.5in}}
\caption{Reconstructed $B$ mass plots for CLEO data
in the channel $B^+ \to \omega\pi^+ + \omega K^+$.
\label{fig:omega}}
\end{figure}

\begin{figure}
\centerline{\psfig{figure=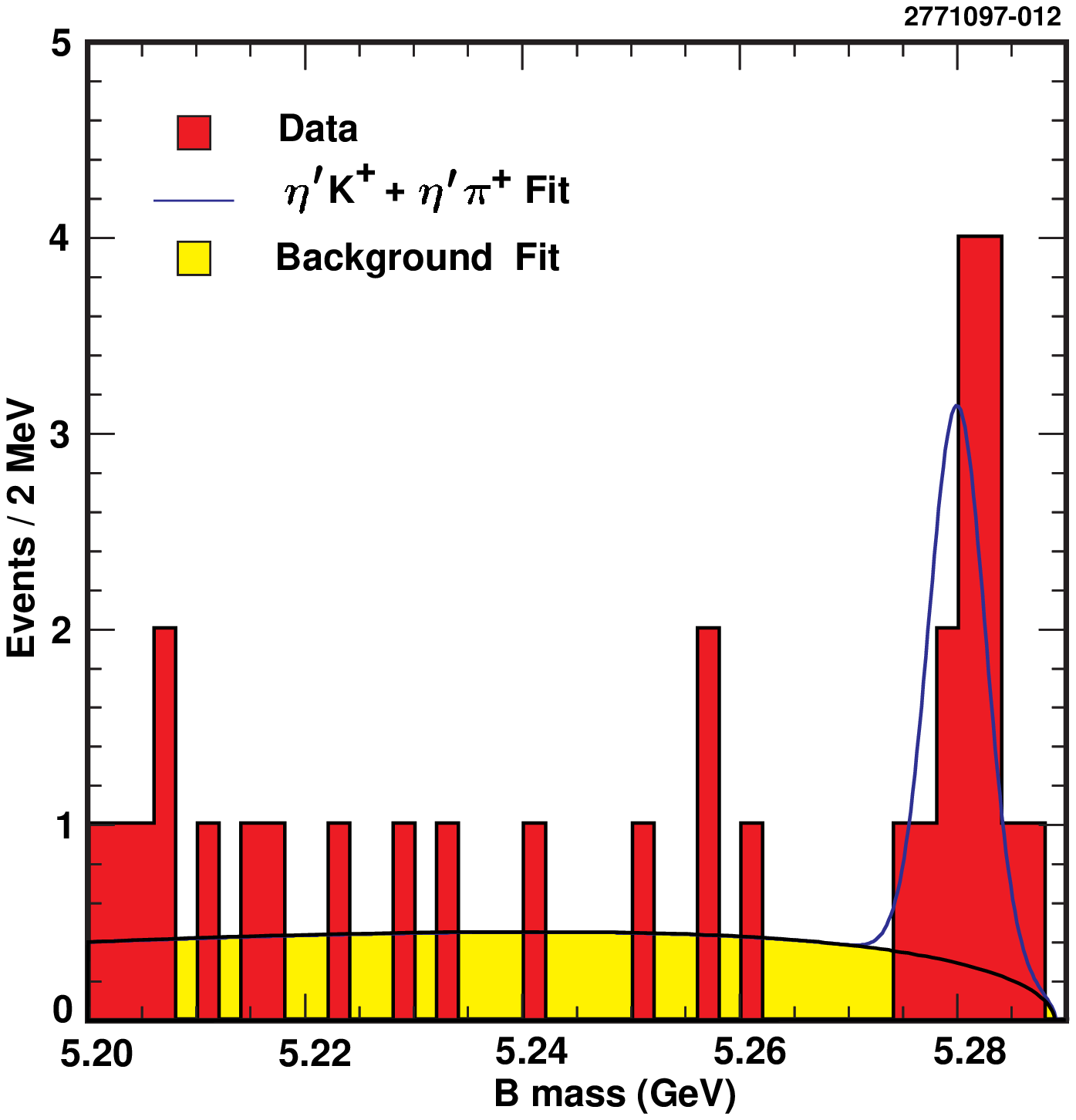,height=2.5in}}
\caption{Reconstructed $B$ mass plots for CLEO data
in the channel $B^+\to \eta^\prime \pi^+ + \eta^\prime K^+$.
\label{fig:eta}}
\end{figure}

The pattern of the hadronic penguin decays of the $B$ meson appearing at a
rate exceeding naive expectation continues.  Signals
are evident in $B^+\to \omega\pi^+ + \omega K^+$ and there is a stunning signal in
$B^+\to\eta^\prime K^+$ as shown in figures ~\ref{fig:omega} and
~\ref{fig:eta}.  The measured branching ratios from
CLEO are still preliminary ~\cite{cleoomega,cleoeta}:
\begin{eqnarray}
{\cal B}(B^+ \rightarrow \omega K^+) & = & \left 
( 1.5 ^{+0.7}_{-0.6} \pm 0.3\right ) \times 10^{-5} \\
{\cal B}(B^+ \rightarrow \omega \pi^+) & = & \left 
( 1.1 ^{+0.6}_{-0.5} \pm 0.2 \right ) \times 10^{-5} \\
{\cal B}(B^+ \rightarrow \eta^\prime K^+) & = & \left
( 7.1 ^{+2.5}_{-2.1} \pm 0.9\right ) \times 10^{-5} \\
{\cal B}(B^+ \rightarrow \eta^\prime \pi^+) & < & 4.5 \times 
10^{-5} ~{\rm ( 90 \% c.l.)}\\
{\cal B}(B^+ \rightarrow \eta h^+ )& < & 0.8 \times 10^{-5} ~{\rm (90 \% c.l.)}
\end{eqnarray}
 The amplitudes for these decays all have many
contributions with the expectation that $B^+ \to \omega K^+$,
$B^+ \to \eta K^+$ and $B^+ \to \eta\prime K^+$ are dominated by the
gluonic penguin diagram.  The $B^+ \to \omega \pi^+$,
$B^+ \to \eta \pi^+$  and $B^+ \to \eta^\prime\pi^+$ modes are
expected to be dominated by
tree diagrams. The rate for the decay $B^+\to\eta^\prime K^+$ is
so large it deserves special note. Even relative to other gluonic penguins such
as $B\to K\pi$ which are larger than expected at the $1.5\times 10^{-5}$ level,
the branching ratio of  $7\times 10^{-5}$ is so significant that many
have speculated that there must be other contributions to the rate than just
the Standard Model penguin. This is further supported by evidence of
anomalously large inclusive branching ratio for 
$B\to\eta^\prime X_s$~\cite{cleoetainc}: 
\begin{eqnarray}
{\cal B}\left( B\to\eta^\prime X_s\right) = \left(6.2\pm1.6\pm1.3\right)\times
10^{-4}; ~ ~~2.0<p_{\eta^\prime}<2.7 {\rm GeV} 
\end{eqnarray}

\begin{figure}
\centerline{\psfig{figure=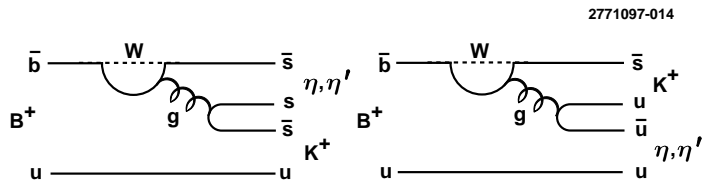,height=1.0in}}
\caption{Two penguin diagrams for the $B^+\to\eta/\eta^\prime K^+$ modes.
Constructive interference between the two diagrams enhances the $\eta^\prime K^+$
final state and destructive interference suppresses the $\eta K^+$ final state.
\label{fig:peng}}
\end{figure}

\begin{figure}
\centerline{\psfig{figure=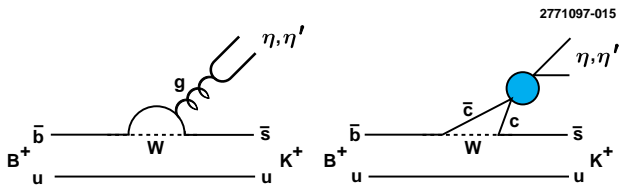,height=1.0in}}
\caption{Two possible explanations for the unexpected enhancement of
$B\to \eta^\prime X$ are illustrated.
\label{fig:etaprime}}
\end{figure}

What could be going on? The most obvious trend is that the gluonic penguins
are larger than we thought. This pattern seems to pervade rare $B$ decays.
The fact that $B^+ \to \eta^\prime K^+$ is so much larger than 
$B^+ \to \eta K^+$ was
not totally unexpected ~\cite{lipkin} and is thought to be due to the interference
of the 2 penguin diagrams shown in figure ~\ref{fig:peng}
with $g\to u\bar u$ and $g \to s\bar s$.
However, the overall enhancement of the $\eta^\prime$ modes has fueled
quite a lot of theoretical speculation ~\cite{explan_eta}, including unexpected enhancement of the
hairpin diagram or a source of $\eta^\prime$ from $b\to c\bar cs$ decays
as schematically illustrated in figure ~\ref{fig:etaprime}. This
latter source of speculation is particularly attractive since 
it helps to 
explain the low value of ${\cal B}\left(B\to X\ell\nu\right)$ which we worried
about earlier. At this moment, speculation is rampant and time and more data
will be needed to sort this out. It is worth pointing out that enhanced gluonic
penguins could be both a blessing and a curse to studies of CP violation in $B$
decays. They can be a source of complication for modes we had hoped were clean.
However, they may also offer more opportunities for studying CP violation in
$B$ decays than previously realized.

\subsection{Effective Flavor Changing Neutral Currents}

I now want to turn to a class of beyond tree level process in heavy quark 
decay that can be described generically as effective flavor changing
neutral current (FCNC) processes.  This
includes decays like $b \to s\gamma, b\to s \ell^+\ell^-, B\to K^{(*)}\ell^+\ell^-$
and $D \to X\ell^+\ell^-$.  These decays are dominated by electromagnetic
and electroweak penguins.  There are no competing tree level processes
as was the case for the gluonic penguins, so these  penguin
and box diagrams offer clean experimental handles for probing
loops in the decays of heavy quarks.
This year saw many new results in charm and bottom decays from the LEP experiments,
CDF and D0, CLEO, E687 and E791.

The effective FCNC decays in charm are very interesting if one sees
a signal, which no one does. 
Unfortunately, the GIM mechanism which is effective in suppressing the
Standard Model rates for loop level charm decays also effectively suppresses
many non-Standard Model rates, and so rare charm decays are not particularly
sensitive as probes of `standard' new physics.  However, one can always
be surprised!   Limits from E687 and E791 are
in the range of $10-20 \times 10^{-5}$  with
Standard Model rates for these decays predicted to be much 
smaller -- typically 
or order $10^{-10}$.

In bottom decays, the experimental sensitivity to loop 
processes is somewhat better
than in charm decays,
depending on the mode. 
Because the GIM
mechanism is not particularly effective in B decays due to the
very large top quark mass,
Standard Model rates are accessible and many non-Standard Model rates are
potentially accessible as well. 
Signals are seen 
in some of the electromagnetic penguin
modes, and both signal rates and upper limits
for FCNC decays
are useful for putting significant model dependent constraints
on extensions of the Standard Model such as supersymmetry.  
The disadvantage of studying FCNC decays in the bottom
system relative to charm is that if one sees a signal,
a good calculation of the expected Standard Model rate is needed.  One 
must look
for deviations from the Standard Model rate to search for evidence of new physics.

The electromagnetic penguin $b \to s \gamma$ has now been seen
 in both exclusive 
 and inclusive
channels.  
The exclusive decay $B \to K^*\gamma$ has been observed by the CLEO experiment ~\cite{cleok*gamma}
with an upper limit placed on the exclusive $B_s\to \phi\gamma$ penguin
decays from
ALEPH ~\cite{alephphigamma}.
\begin{eqnarray}
{\cal B}(B \to K^*\gamma) & = & \left ( 4.2 \pm 0.8\pm 0.6 \right ) \times 10^{-5} {\rm ~(CLEO)}\\
{\cal B}(B_s \to \phi\gamma) & < &  29 \times 10^{-5} {\rm ~~90\% c. l.  ~(ALEPH)} 
\end{eqnarray}
 
\begin{figure}
\centerline{\psfig{figure=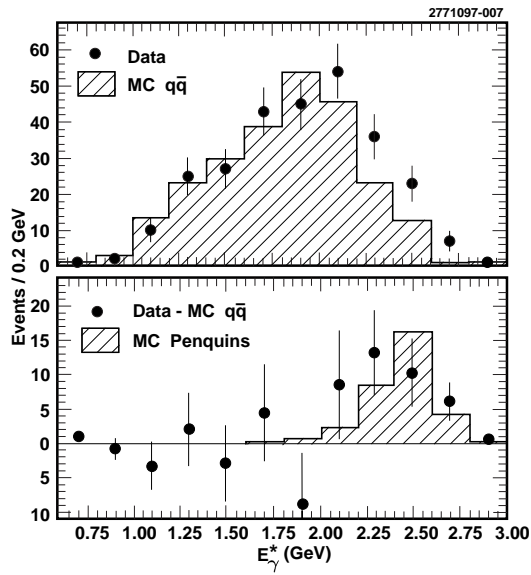,height=3.0in}}
\caption{The photon spectrum showing the signal for $b\to s \gamma$ from ALEPH.
The top plot shows the raw photon spectrum.  The points are data and the histogram
is Monte Carlo expectation with no penguin contribution.  In the bottom plot the
Monte Carlo photon spectrum for photons from all other sources has been subtracted
showing the signal for the penguin decay.  The points are again data and the histogram
is the expectation of the photon spectrum for $b\to s \gamma$ decays from Monte Carlo.
\label{fig:bsgamma}}
\end{figure}

The more interesting channel is
the inclusive electromagnetic penguin  where calculations are thought to be
much more reliable. 
CLEO first observed the inclusive electromagnetic penguin decay~\cite{cleobsgamma}
and ALEPH has recently reported a preliminary 
observation of this decay~\cite{alephbsgamma}.
\begin{eqnarray}
{\cal B}(B \to s\gamma) & = & \left ( 2.32 \pm 0.57\pm 0.35 \right ) \times 10^{-4} {\rm ~(CLEO)}\\
{\cal B}(B \to s\gamma) & = & \left ( 3.38 \pm 0.74\pm 0.85 \right ) \times 10^{-4} {\rm ~(ALEPH)}
\end{eqnarray}
The ALEPH analysis takes advantage of the
long flight of the $ B$ meson at LEP to tag a $ b$ jet in one hemisphere
and then requires a high energy photon in the opposite hemisphere.
They use rapidity, momentum and impact parameter to
discriminate tracks produced in the $b$ decay from tracks from primary hadronization.
In this way they inclusively reconstruct the candidate $B$ meson mass.
After assorted  cuts, they fit the photon energy distribution in the $B$ candidate
rest frame to extract the $b \to s\gamma$ rate.  The photon spectrum from the ALEPH
analysis is shown in figure ~\ref{fig:bsgamma}.

Dramatic progress has been made in the theoretical interpretation
of the inclusive electromagnetic penguin decays where two new
calculations of the $b \to s \gamma$ rate now include all terms to
next-to-leading order ~\cite{chetyrkin,buras}.  
The new calculations give a slightly  increased  theoretical
Standard Model rate
for the process and substantially reduce the theoretical errors.  This
results in tighter (although still model dependent) constraints on
new physics from this decay rate.

The other particularly interesting FCNC decays are the 
decays with a lepton pair in the final state.  No one has yet seen a signal.  
The recent results on these
decays are summarized in table ~\ref{tab:fcnc}. The interesting point to note here
is that a wide variety of experiments are now within an order
of magnitude of the Standard Model expectation for these decays.

\begin{table}[t]
\caption{Summary of the recent results on effective flavor
changing neutral currents in $B$ and $D$ decay.  The 
experimental results on the branching ratios are given
along with the theoretical expectation in the Standard
Model.
\label{tab:fcnc}}
\vspace{0.4cm}
\begin{center}
\begin{tabular}{|c|c|c|c|}
\hline
 & & & \\
Mode & ${\cal B}_{expt}$ & Expt.  & ${\cal B}_{th}$  \\
      & $(10^{-5})$ & &  $(10^{-5})$ \\
& & & \\
\hline
&  & & \\
$b\to s\gamma$ & $25.5 \pm 6.1$ & CLEO/ALEPH 
~\cite{cleobsgamma,alephbsgamma} & $34 \pm 3$ ~\cite{chetyrkin,buras}\\
$B \to K e^+e^-$  &  $<1.2$ &CLEO~\cite{cleoKll} & 0.02 - 0.05 ~\cite{Kllth}\\
$B \to K \mu^+\mu^-$  &  $<0.9$ &CLEO~\cite{cleoKll} & 0.02 - 0.05
~\cite{Kllth} \\
$B\to K^* e^+ e^-$ & $<1.6$ & CLEO~\cite{cleoKll} & 0.2 - 0.5~\cite{Kllth} \\
$B\to K^* \mu^+\mu^-$ & $<2.5$ & CDF~\cite{cdfKll} & 0.2 - 0.5~\cite{Kllth} \\
$b \to s\mu^+\mu^-$  & $<5.8$ & CLEO~\cite{cleosll}  & 0.6~\cite{sllth} \\
$b \to s e^+ e^-$ & $<5.7$ & CLEO~\cite{cleosll} & 0.8~\cite{sllth} \\
$b \to s \nu \bar\nu$ & $<77$  & ALEPH~\cite{alsnn} & 4~\cite{snnth}\\
$B_d \to \mu^+ \mu^-$ & $ <0.026 $ & CDF~\cite{cdfmm}& 
$\sim 0.00001$ ~\cite{Kllth} \\
$B_s \to \mu^+ \mu^-$ & $ <0.077 $ & CDF~\cite{cdfmm}&
$\sim 0.0001$  ~\cite{Kllth}\\
$D^+ \to \pi^+ e^+ e^-$ & $<6.6$ & E791~\cite{e791rare} & 
$<0.001$ ~\cite{hewett}\\
$D^+ \to \pi^+ \mu^+ \mu^-$ & $ <1.8 $ & E791~\cite{e791rare} &
  $<0.001$ ~\cite{hewett} \\
$D^+ \to K^+ e^+ e^-$ & $<20$ & E687~\cite{e687rare} & $<10^{-10}$ ~\cite{hewett} \\
$D^+ \to K^+ \mu^+ \mu^-$ & $<9.7$ & E687~\cite{e687rare} & 
$<10^{-10}$  ~\cite{hewett}\\
& & &\\
 \hline
\end{tabular}
\end{center}
\end{table}

\begin{figure}
\centerline{\psfig{figure=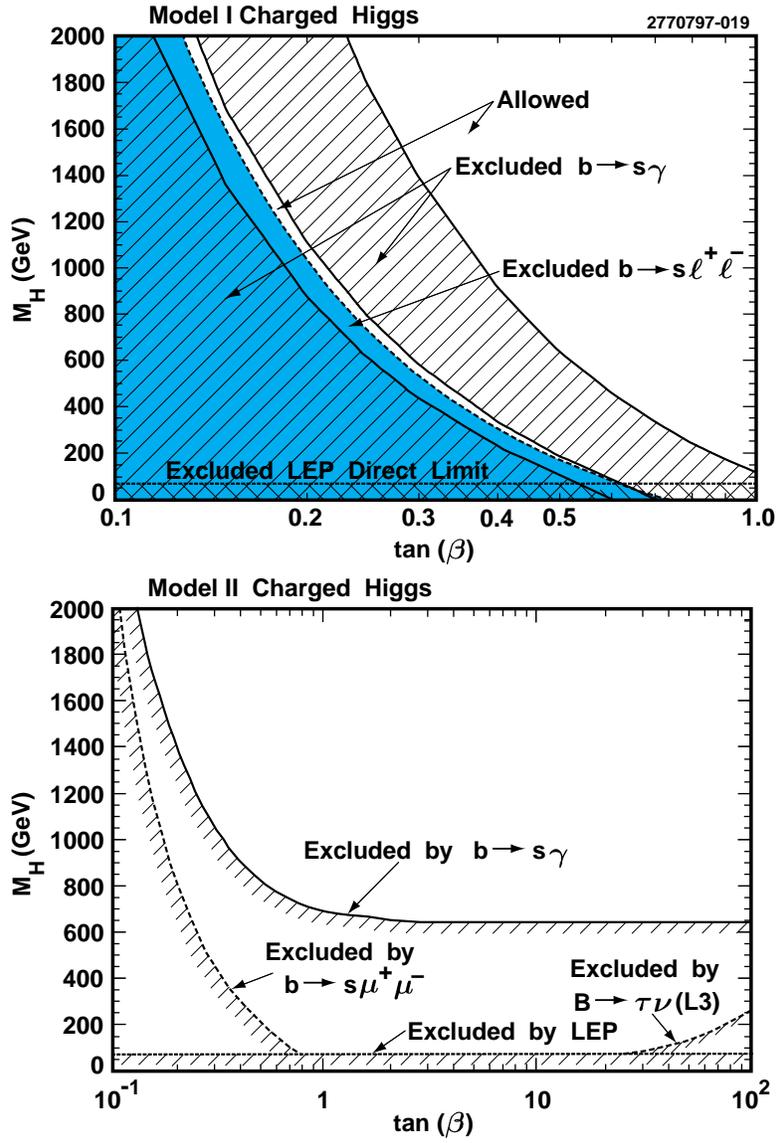,height=6.0in}}
\caption{The parameter space for two generic models with extra Higgs is shown
with the areas excluded by the FCNC results shaded.  The LEP direct limit
is also shown.
\label{fig:fcnc}}
\end{figure}

Figure ~\ref{fig:fcnc} illustrates the power of the effective FCNC decays.
The shaded areas are the regions of parameter space that are
excluded for two generic types of models that have
extra Higgs.  SUSY is an example of type II, but the
limits shown here assume that the charged Higgs and nothing else
is affecting the $b\to s \gamma$ rate.  The limits are powerful,
but they are also (with the exception of the LEP direct limits)
model dependent ~\cite{higgstype}.

\begin{figure}
\centerline{\psfig{figure=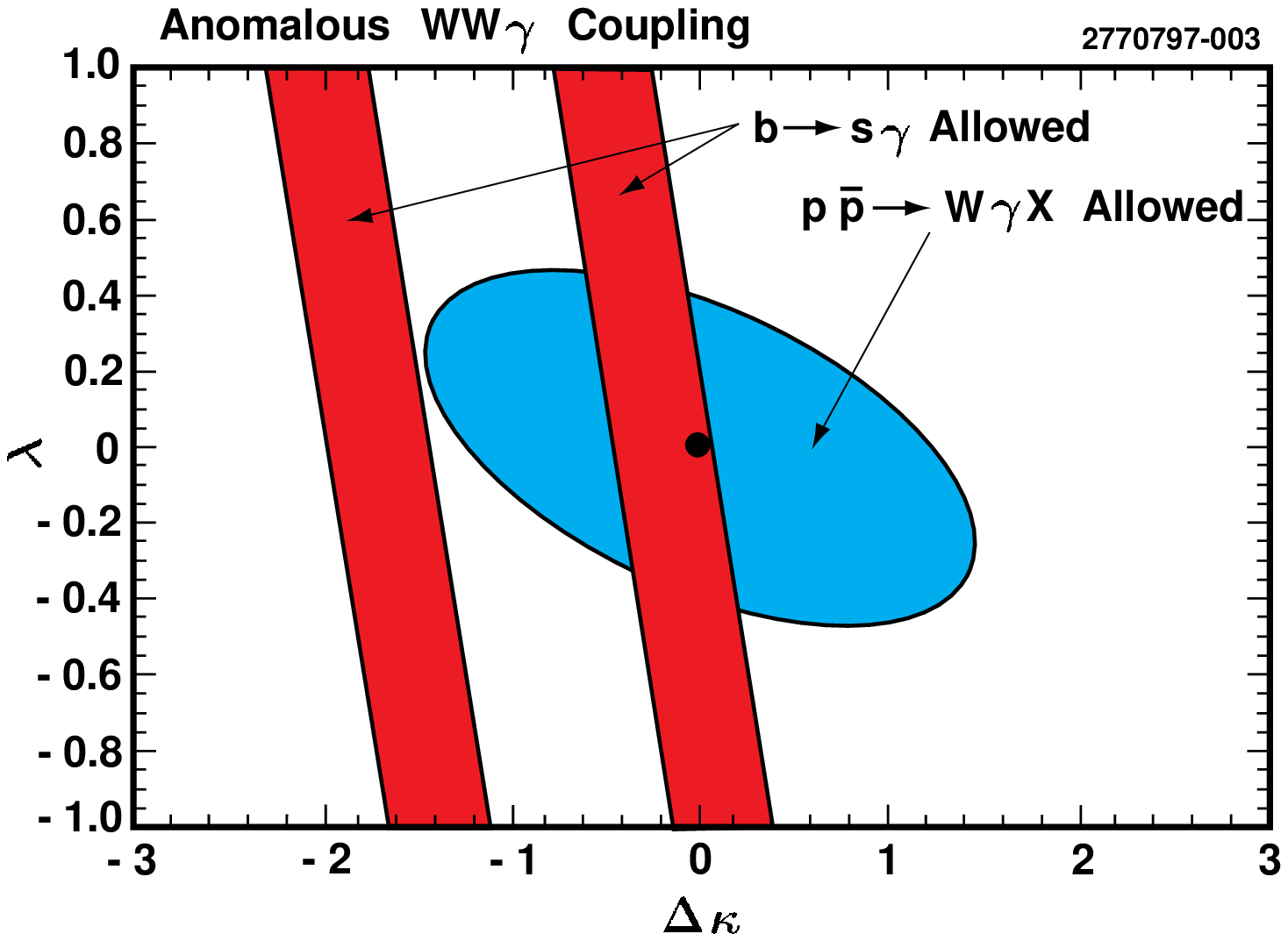,height=2.8in}}
\caption{The allowed values for  $WW\gamma$ anomalous couplings from
the measured $b \to s \gamma$ branching ratio and the $D0$ result
on $p\bar{p} \to W \gamma X$ are shown as shaded regions.
\label{fig:anom}}
\end{figure}

Another  way to present the constraints  from the FCNC decays is
to show how they restrict the $WW\gamma$ anomalous couplings.  In figure
~\ref{fig:anom}, the
region allowed by $b\to s \gamma$ cuts an impressive swath across the
region allowed by $D0$ from its analysis of $p\bar p \to W\gamma X$ ~\cite{higgstype}.

\section{Conclusions}
Summarizing the status of heavy quark decays, the tree level processes
are reasonably well understood, but we need to understand
them even better if we want to
improve our knowledge of $|V_{ub}|$ and $|V_{cb}|$.  More 
data in both charm and bottom decays will be necessary to accomplish
this.  There is no strong evidence of any serious misunderstanding 
in the tree level decays.  There are a variety of 2$\sigma$ problems
that either need to become 3$\sigma$ problems or go away!  

The field of heavy quark decays beyond tree level is emerging and rapidly
developing.  There have been many surprises in the
hadronic rare $B$ decays.  The gluonic penguins seem to 
be consistently larger than expected. The
FCNC decays are providing substantial (although model dependent)
constraints on new physics.  However, the goal of
studying the phases of elements of the CKM matrix is still a ways
off.  I think that we can eagerly look forward to the Lepton--Photon
Conference perhaps 4 years from now when data will be starting to
probe the CKM phases.  We are at the beginning of a new era in heavy
quark physics with studies of rare processes.  I look forward
to exploring where it leads us.

\section*{Acknowledgments}
Many people were generous with their time and effort in helping me to prepare
this review.  I am particularly indebted to
David Cassel who provided the branching ratio averages for the
semileptonic $B$ decays.  Lawrence Gibbons and
Ken Bloom graciously supplied
the new $|V_{cb}|$ averages.  Claudia
Glasman was a wonderful scientific secretary.
I am grateful to Veronique Boisvert, David Crowcroft, and
Andy Foland who read the manuscript carefully and
made many thoughtful comments.
This work was supported by the National Science
Foundation.

\end{document}